\documentclass[12pt,preprint]{emulateapj}

\newcommand{\simgt}{\,\rlap{\lower 3.5 pt \hbox{$\mathchar \sim$}} \raise
1pt \hbox {$>$}\,}
\newcommand{\simlt}{\,\rlap{\lower 3.5 pt \hbox{$\mathchar \sim$}} \raise
1pt \hbox {$<$}\,}
\newcommand{\vrot}{v_{\textrm{rot}}}

\shorttitle{DM angular momentum profile}
\shortauthors{Schmidt et al.}

\begin{document}


\title{Dark matter angular momentum profile from the Jeans equation}


\author{Kasper B. Schmidt$^{\dag \; \star}$, Steen H. Hansen$^\dag$, Jin H. An$^\dag$, Liliya  L. R. Williams$^\ddag$, \& Andrea V. Macci\`o$^\star$
}
\affil{$^\dag$ Dark Cosmology Centre, Niels Bohr Institute, University of Copenhagen,\\
Juliane Maries Vej 30, 2100 Copenhagen, Denmark}
\affil{$^\ddag$ Astronomy Department, University of Minnesota,
116 Church Street SE, Minneapolis MN 55455, USA}
\affil{$^\star$ Max Planck Institut f\"ur Astronomie, K\"onigstuhl 17, 69117 Heidelberg, Germany}



\begin{abstract}
Cosmological simulations of dark matter structures have shown that the
equilibrated dark matter structures have a fairly small angular
momentum. 
It appears from these N-body simulations that the radial profile of
the angular momentum has an almost universal behavior, even if the
different dark matter structures have experienced very different
formation and merger histories. We suggest a
perturbed Jeans equation, which includes a rotational term.  This is
done under a reasonable assumed form of the change in the distribution
function.  By conjecturing that the (new) subdominant rotation term must be
proportional to the (old) dominant mass term, we find a
clear connection, which is in rather good agreement with the results
of recent high resolution simulations. We
also present a new connection between the
radial profiles of the angular momentum and the velocity anisotropy,
which is also in fair agreement with numerical findings. 
Finally we show how the spin parameter $\lambda$ increases as a function of radius.
\end{abstract}




\section{Introduction}
\label{sec:intro}

Our understanding of dark matter structures has increased
significantly over the last years. This progress has mainly been
driven by pure dark matter numerical simulations which have suggested
or identified a range of universalities. One of the first general
properties to be suggested is the radial density profile
\citep{nfw,moore,merritt,alister}.  Also more complex connections
relating integrated quantities have been suggested, including the
pseudo phase-space density being a power-law in radius
\citep{taylornavarro}, $\rho/\sigma^3 \sim r^{- \alpha }$, or a
connection between the velocity profile and the density slope
\citep{hansenmoore,hansenstadel}, $\beta \sim \gamma$
, where $\gamma = \frac{d\ln \rho}{d\ln r}$ and $\beta = 1-\frac{\sigma_t^2}{\sigma_r^2}$.
Finally, a connection between angular momentum and mass has been identified
\citep{bullock}, $j \sim M^s$. Only few attempts have been made at
identifying universalities in the actual velocity distribution function
~\citep{hansenzemp,wojtak}, since non-integrated quantities require a very
large number of particles in the equilibrated structure.

A wide range of theoretical ideas and models have been devised trying
to explain these phenomenological profiles and relations. First of
all, the general properties of the density profiles~\citep{nfw,moore}
can be derived analytically under the assumption that phase-space
density is a power-law in radius
\citep{taylornavarro,hansenjeans,austin,dehnenmclaughlin}. This is done simply by
inserting the phase-space density into the Jeans equation, and then
solving it.
This procedure reveals one unique mathematical solution, which is physically plausible in the equilibrated region of the dark matter structures.

A completely different approach is made in the Barcelona
model~\citep{ssm98,mrsss03,smgh07}, where slow accretion in a
generalized Press-Schechter model allows one to derive density
profiles which are in very good agreement with the profiles observed
in numerical simulation.  

It thus appears that there are (at least)
two completely different possible explanations for the structural
property of the density: one being that the density profile is slowly
grown according to the parameters in the expanding universe, and the
other possibility is that irrespective of how the structures are
formed, then some unknown dynamical process forces the phase-space density to
be a power-law in radius, which through the Jeans equation gives the
density profile itself. \cite{austin} argue that this process is violent relaxation. A different theoretical approach to get the
density profile is a series expansion of the coupled collisionless
Boltzmann  and Poisson equations which is renormalized in
`time' \citep{henriksen}. However, when comparing to numerical results,
it appears that this last theoretical approach still leaves room for
improvements~\citep{henriksen}.

The relation between angular momentum and mass, which was discovered
in numerical simulations~\citep{bullock}, has been explored in the
Barcelona model of accretion-driven formation of cosmological
structures \citep{gsmh}. It appears that the results of this
theoretical model are in rather good agreement with the results of
numerical simulations. It is therefore tempting to conclude, that the
kinematic and structural parameters may indeed be governed by the
physics behind the Barcelona model - at least until we find another
convincing derivation of the $j-M$ relation based on different
assumptions. This is exactly what we set out to do in this paper. 

In general when dealing with the angular momentum of dark matter in galaxies people often use the so-called spin parameter $\lambda \equiv \frac{L |E|^{1/2}}{GM^{5/2}}$, introduced by \cite{peebles69}, as a common reference point. Here $L$, $E$, $M$ and $G$ are the angular momentum, the binding energy and the total mass of the system and the Newtonian gravitational constant respectively. The spin parameter roughly corresponds to the ratio between the overall angular momentum of the object and the angular momentum this object needs to sustain rotational support. One of the reasons why $\lambda$ is so widely accepted as a good indicator of the angular momentum, is that it has the ability of being (almost) constant in time, under the assumptions that the system  is more or less isolated  and that there is no dissipation present (assuring that $E$ and $L$ are both conserved). The usual size of the spin parameter is in the ballpark of 0.05 \citep{vitvitska02}, meaning little systematic rotation and negligible rotational support.

There is still no agreement on the origin of the angular momentum.
Two different scenarios have been tested against each other (e.g. by \cite{maller02}). The first scenario states that the angular momentum of structures originates from the merging history, and hence is dependent on how and when the given structure formed \citep{vitvitska02}. The second scenario explains the appearance of angular momentum as a consequence of linear tidal torques between density fluctuations in the early stages of galaxy formation \citep{dongnavarro}.

We will show below that the generalized Jeans equation includes a term
governing the profile of the angular momentum. This allows us to
suggest an angular momentum profile directly from the Jeans equation,
with no reference to the specific way the structures were
assembled. This contrasts the claim that merger history is crucial when describing angular momentum. We use results from recent high resolution N-body
simulations to show that the angular momentum profile of the
numerically equilibrated dark matter (DM) structures shows good agreement with our
suggested relation. This indicates that the angular momentum
profile is fixed through the Jeans equation, irrespective of the
detailed structure formation process.
This result is supported by the numerical results of \cite{ascasibar} and \cite{dongnavarro}, who conclude that "equilibrium dark matter haloes show
no significant correlation between spin and merging history".

We present a connection between
the radial profiles of the angular momentum and the velocity anisotropy.
This correlation is also shown to be in fair agreement with numerical simulations.

Finally we combine our suggested relation between the angular momentum and the mass with the spin parameter, and find that the radial dependence of this spin parameter is in fair agreement with the results of \cite{ascasibar}.


\section{Jeans equation including rotation}
\label{sec:jeans}

In general nothing ensures that the particle ensemble in a DM halo is spherical. Nevertheless it makes the approach relatively simple and analytically manageable, without being far from the 'real' triaxial picture. The system is then governed by the spherical(ly symmetric) collisionless Boltzmann equation (CBE) (with the velocity distribution function being a function of the radial dependence alone). One might also use other (more complicated) forms of the DF (e.g. \cite{tonini2006}) but this would make the approach analytically cumbersome. The CBE describes the relation
between the spatial distribution function, the gravitational potential and the
velocity distribution of the DM particles. 

Combining the first moment of the CBE with the spherical Poisson
equation, one can write the general spherical Jeans
equation for the DM particle ensemble, 
under the assumption that the thermal velocities are independent
of each other, and
that there is no bulk (rotational) motion~\citep{binneytremaine}
\begin{equation}\label{eqn:CJE}
\sigma_r^2  \frac{d\ln (\rho \sigma_r^2)}{d\ln r} + \left(2 \sigma_r^2 - \sigma_{\theta}^2  - \sigma_{\phi}^2  \right) = - \frac{GM}{r} \, .
\end{equation}
Here $\sigma_{\theta}^2$, $\sigma_r^2$ and $\sigma_{\phi}^2$ are the
velocity dispersions, $\rho$ is the density, $M$ is the total mass 
within radius $r$, and $G$ is the gravitational constant.

We wish to consider the angular momentum, so we will now include a
small radial dependent bulk rotation, $\vrot (r)$.  
We explicitly write the velocity in the following way
\begin{equation}
v_{\phi} (\vrot = 0) \rightarrow \tilde{v}_{\phi} (\vrot) \, ,
\end{equation}
which means that the azimuthal velocity goes from being
purely thermal to containing both thermal and bulk motion.

In general $\vrot$ is not a function of the spherical radius $r$, but more likely a function of the cylindrical radius $R$. So when we in the following refer to $\vrot(r)$, what we really mean is, that the radial velocity has been averaged over a spherical shell with radius $r$. This is also the way $\vrot$ is defined in the simulations we will compare with in section~\ref{sec:sim}. Furthermore since the effect of bulk rotation cannot depend on direction, we only consider the absolute value of $\vrot$.

The simplest way to add rotation to our system, is to give every particle an initial kick. This would result in  a shift of the distribution function (DF) $f$, towards a higher mean velocity by the amount added, call it $\vrot$. Since we need the structure to be a relaxed system in equilibrium we can't allow this. A pure shift of the DF would enable the most energetic particles to escape the system, and thereby bring it out of equilibrium. Instead of just shifting the DF we force it to be distorted into a new DF $\tilde{f}$, making sure that the system is always in equilibrium.

To be able to implement such a distortion into Eq.~(\ref{eqn:CJE}) we will now parametrize the distortion of the DF. As a first step, we assume that the difference in the two DFs is just a small perturbation to the overall system. Furthermore the distortion must depend on the added bulk rotation so that we have
\begin{equation}
f-\tilde{f} \propto P(\vrot) f	\; .
\end{equation}
Here $P$ is some unknown function depending on $\vrot$ which we will evaluate later.

As mentioned above, simply shifting the DF by $v_{\textrm{rot}}$ while leaving the shape of the DF virtually unchanged would allow energetic particles to escape. 
The perturbation of the DF must therefore also depend on the azimuthal velocity itself, such that the amount of distortion is not the same at all velocities. Introducing another unknown function $Q$ we then have
\begin{equation}\label{eqn:ffa}
f-\tilde{f} \propto P(\vrot)\; Q(v_\phi) f
\end{equation}

Furthermore, we will demand that the overall
density of the system  is not affected by the added bulk rotation, i.e, $\rho =
\tilde{\rho}$. The density is statistically defined as
 \begin{eqnarray}
 \tilde{\rho} \;= \int \tilde{f}d^3v 
&=& \int f-(f-\tilde{f}) d^3v = \rho - \int (f-\tilde{f})d^3v \; \quad
 \end{eqnarray}
where we have used that $\tilde{f} = f-(f-\tilde{f})$.
Ensuring that the density is unaffected by the added rotation is easily done by restricting $(f-\tilde{f})$ to be an odd function when integrated over the velocities, such that the last integral vanishes. Assuming that $Q$ is a simple power law in the azimuthal velocity with a positive odd integer exponent (to be motivated later in this section) this is accomplished and we have
\begin{equation}\label{eqn:ff}
f-\tilde{f} = \xi P(\vrot) \left( \frac{v_{\phi}}{(\sigma_{\phi}^2)^{1/2}} \right)^{\gamma}  f
\end{equation}
with $\gamma = 2n+1$ for $n=0,1,2,...$ and $\xi$ being an unknown constant. 

The $Q(v_\phi)$ part of
the parameterization on this power law form will in principle give problems for $v_{\phi} \rightarrow
-\infty$, causing the distortion of the DF to become infinite and hence $\tilde{f}$ to become negative, but since $f(v_{\phi})$ is nearly zero for $v_{\phi} \sim
\pm 4\sqrt{\sigma_{\phi}^2}$, this is not a practical problem. However, for a more formal derivation one naturally  cannot allow $\tilde{f}$ to become negative.

Thus, by making $\gamma$ an odd integer, what we have done (using Eq.~(\ref{eqn:ffa}))  is to create a new DF, $\tilde{f}$, that generates the density $\rho$ by adding an odd-powered DF concerning $v_\phi$ to our original DF, $f$.
This is somewhat similar to the discussion in \cite{binneytremaine} section 4.5. 

In principle, the assumption in Eq.~(\ref{eqn:ff}) can be tested with
high resolution N-body simulations, by considering the two tangential
velocity distribution functions (VDFs), and simply looking for the
difference between the shapes of the VDFs in the directions parallel
and perpendicular to the angular momentum vector. This we can do using the N-body/gas dynamical simulated large disc galaxy 'K15'  \citep{jsl,hansenjsl}.
The 'K15' simulation is a significantly improved version of the TreeSPH code used previously for galaxy formation simulations \citep{jslgp}. This simulation is  based on a flat $\Lambda$CDM model with $(\Omega_M,\Omega_\Lambda)=(0.3,0.7)$. The simulation takes many of the important factors of galaxy formation into account such as SN feedback, gas recycling (tracing 10 elements), atomic radiative cooling, etc. It consist of both cold and warm gas, DM, disk and bulge stars and stellar satellites. The galaxy 'K15' contains about $3\times10^5$ gas and DM particles and it has $m_\textrm{gas}=m_\textrm{stars}=7.3\times10^5 M_{\odot}/h$ and $m_\textrm{DM}=4.2\times10^6 M_{\odot}/h$ where $h=0.65$. Furthermore the gravitational (spline) softening lengths adopted are $\epsilon_\textrm{gas}=\epsilon_\textrm{stars}=380$ and $\epsilon_\textrm{DM}=680 \; \textrm{pc}/h$.

\begin{figure}
 \epsscale{0.95}
 \plotone{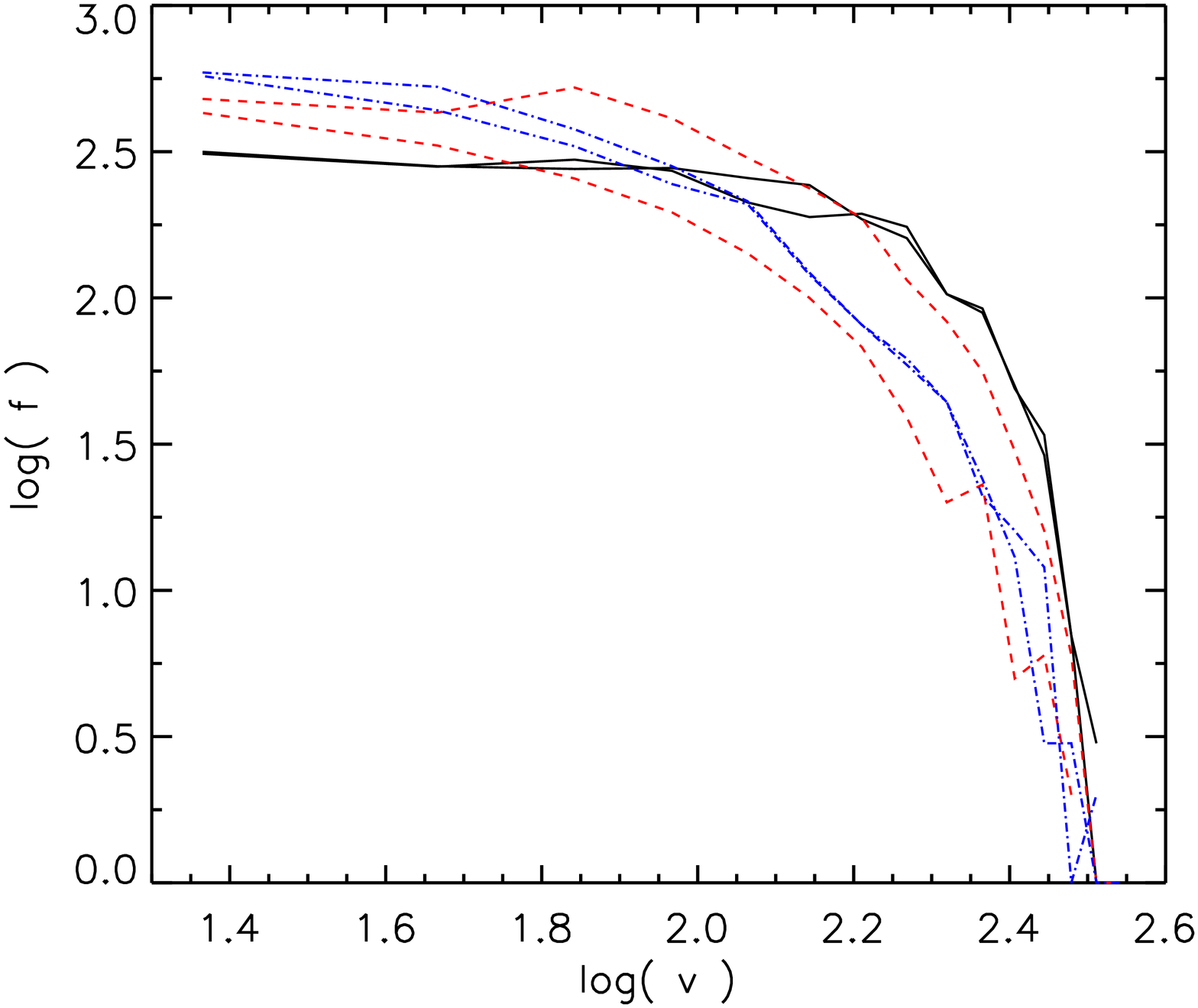}
	\caption{ The velocity DFs for the DM particles in one of the potential bins in the galaxy K15 in the radial direction (black solid line) and the tangential direction perpendicular to (red dashed line) and parallel to (blue dot-dashed line) the angular momentum vector for the considered bin. We clearly see that the DF containing rotation (perpendicular to the angular momentum vector) is distorted as suggested. All the trustable bins in 'K15' show a similar picture. This supports our assumption of the added rotation causing a distortion of the DF of the DM structure.}
\label{fig:dfdis}
\end{figure}

In Fig.~\ref{fig:dfdis} we have plotted the VDFs of the DM particles of one of the bins outside the stellar disk region in 'K15'. We clearly see that the DF including rotation (red dashed line) is indeed distorted compared to the radial VDF (black solid line) and the tangential VDF parallel to the angular momentum vector (blue dot-dashed line). This supports our assumption of a distortion of the DF caused by the added bulk rotation. Furthermore we realize that the distortion seems to grow (i.e., the difference between the two red dashed lines enhances) as the velocity grows out to about 2-3 $\sigma$, beyond which the number of particles in each bin of the simulation is too small to conclude anything. This further motivates a form of $Q$ similar to the one suggested in Eq.(\ref{eqn:ff}).

To be able to see the effects the added rotation has on the system in general, the goal is to get an expression for the Jeans equation concerning a system to which a small bulk rotation has been added. 
This can be accomplished by making an expression for the new perturbed velocity dispersion in the azimuthal direction, $\tilde{\sigma}_{\phi}^2$. The velocity dispersion of a system without rotation is given by
\begin{equation}\label{eq:dispdef}
\sigma_{\phi}^2 = \frac{1}{\rho} \int f  v_{\phi}^2  \; d^3v \; .
\end{equation}
By definition the velocity dispersion is the integral over the DF multiplied with the difference between the individual particle velocities and the mean velocity of the system. Since the mean velocity after adding rotation becomes equal to the added rotation itself, the new perturbed azimuthal velocity dispersion must take the form
\begin{equation}\label{eq:dispdef2}
\tilde{\sigma}_{\phi}^2 = \frac{1}{\rho} \int \tilde{f}  \left(v_{\phi}-\vrot\right)^2  \; d^3v 
\end{equation}
where it is easily shown that $\vrot = 1/\rho \int \tilde{f} v_\phi d^3v$.

Using that $\int f d^3v = \rho$, i.e., $f$ is an even function in velocity space, that $(f-\tilde{f})$ is an odd function to conserve the density and that $\tilde{f} = f-(f-\tilde{f})$, combining Eqs.~(\ref{eq:dispdef}) and (\ref{eq:dispdef2}) gives
\begin{equation}
\tilde{\sigma}_\phi^2 = \frac{1}{\rho} \int f v_{\phi}^2  \;  d^3v +\frac{\vrot^2}{\rho} \int f  \;  d^3v + \frac{2\vrot}{\rho} \int (f-\tilde{f}) v_{\phi}  \;  d^3v	\; .
\end{equation}
Since it is known from numerical cosmological simulations that the rotational energy is less than a few percent of the thermal energy \citep{bullock}, i.e., $\vrot^2 \ll \sigma_\phi^2$ it is justified to ignore  the higher order term in $\vrot$. This combined with Eq.~(\ref{eq:dispdef}) implies that
\begin{equation}\label{eq:vphi}
\tilde{\sigma}_\phi^2 \approx \sigma_\phi^2  + \frac{2\vrot}{\rho}\int (f-\tilde{f}) v_\phi \; d^3v	\; .
\end{equation}
Thus the new velocity dispersion can be written as the old one plus a term concerning the distortion of the DF as well as the rotation.
Combining this with the DF distortion in Eq.~(\ref{eqn:ff}) implies
\begin{equation}\label{eq:vphi2}
\tilde{\sigma}_\phi^2 = \sigma_\phi^2 +  2\xi\vrot P(\vrot) (\sigma_\phi^2)^{-\frac{\gamma}{2}} \frac{1}{\rho} \int v_\phi^\gamma v_\phi \, f \; d^3v	\; .
\end{equation}
We are then left with evaluating an integral on the form
\begin{equation}\label{eq:momint}
 \frac{1}{\rho} \int v_\phi^{k}f \; d^3v
\end{equation}
where $k=\gamma+1$ is an even number (gamma is odd).

Recognizing that the integral in Eq.~(\ref{eq:momint}) is just the expression for the $k$'th moment, $\alpha_k$, for $v_\phi$ centered around the mean $\langle v_\phi \rangle =0$ gives
\begin{equation}\label{eq:moment}
 \frac{1}{\rho} \int v_\phi^{k}f \; d^3v = \alpha_{k}(\sigma_\phi^2)^{k/2} \; \textrm{with} \; k=2n+2 \; \textrm{for} \; n=0,1,2,...
\end{equation} 
Combining this with the expression for the perturbed velocity dispersion in Eq.~(\ref{eq:vphi2}) gives
\begin{equation}\label{eq:sig0}
\tilde{\sigma}_\phi^2 = \sigma_\phi^2 + 2\, \xi \, \vrot \,  \alpha_{\gamma+1} \,  P(\vrot) \,  \sqrt{\sigma_\phi^2} \;.
\end{equation}

\begin{figure}
 \epsscale{0.95}
 \plotone{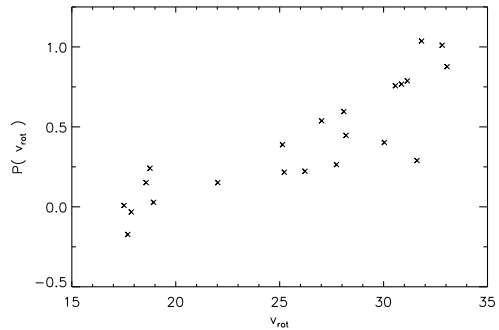}
	\caption{ The function $P(\vrot)$ from the simulation 'K15' \citep{jsl,hansenjsl} plotted as a function of $\vrot$ in the trustable region, i.e., from around 10 kpc (where most of the gas disk vanishes) and out to the virial radius. For $\vrot\rightarrow 0$ one must have $P \rightarrow 0$ which however is outside the numerically resolved region in this specific case.}
\label{fig:pvrot}
\end{figure}

We are now able to quantify the $\vrot$ dependency of the distortion of the DF, i.e., the function $P$. An easy way to evaluate the function $P$ is by rearranging Eq.~(\ref{eq:sig0}) so that
\begin{equation}
P(\vrot) \sim \frac{\tilde{\sigma}_\phi^2 - \sigma_\phi^2}{\vrot \sqrt{\sigma_\phi^2}}	\; . 
\end{equation}
Using the 'K15' simulation again we are then able to estimate the actual size of $P$ as a function of the rotation. 
Calculating the relevant quantities from the simulation and plotting $P$ as a function of $\vrot$ gives Fig.~\ref{fig:pvrot}. From this figure, which is showing the numerically resolved region of the structure, we see that $P$ seems to depend liniarly on the rotation, meaning that in this region of the structure $P=C_1+C_2 \vrot $ where the $C$s are constants. However one must keep in mind that (outside the resolved region) $P$ must go to 0 for vanishing $\vrot$. Combining this with expression (\ref{eq:sig0}) and again ignoring higher order terms in $\vrot$ leaves us with
\begin{equation}\label{eq:sig1}
\tilde{\sigma}_\phi^2 = \sigma_\phi^2 +   \eta \, 12 \, \alpha_{\gamma+1} \, \vrot \,  \sqrt{\sigma_\phi^2} \; .
\end{equation}
Here we have introduced the constant $\eta=\frac{C_1\xi}{6}$ (for simplicity). Thus, what we have done here is simply inserting the linear dependence of $P$ on $\vrot$ into Eq.~(\ref{eq:sig0}) and using that  $\vrot^2 \ll \sigma_\phi^2$ as done in Eq.~(\ref{eq:vphi}).

As mentioned $\alpha$ is just the moment corresponding to the chosen value of $\gamma$ (where $Q\sim v_\phi^\gamma$), i.e.,  a constant. And since the above result only relies on the restrictions on $\gamma$ it implies that the choice of $\gamma$ doesn't result in loss of generality. Thus we are free to choose any value of $\gamma$ when investigating the above equation. We will chose $\gamma=3$ since we are then able to estimate the size of the moment. The fourth moment, i.e.,  the kurtosis of a Gaussian DF is 3, and since we expect the DFs of DM structures to be Gaussian like, using $\gamma=3$ will make it easier to compare with simulations. This implies that
\begin{equation}\label{eqn:sig}
\tilde{\sigma}_{\phi}^2  = \sigma_{\phi}^2 + \eta \, 12 \, \alpha_4 \,
v_{\textrm{rot}} \sqrt{ \sigma_{\phi}^2}  \, .
\end{equation}
This expression is of course a consequence of the assumed form of $Q$. One could definitely argue for other forms of $Q$ fulfilling the request $\rho=\tilde{\rho}$, e.g. an exponential form or a combination of both exponential and power law. However in order to simplify the analytical calculation of the integral in Eq.~(\ref{eq:vphi}) we have chosen the simple power law form. 
In the future it would definitely be interesting to test other forms of $Q$ (and $P$) to see if this effects the final results and conclusions significantly.
We intend to make a qualitative estimation of both Q and P in a following paper. This will, among other things, require a larger sample of high resolution equilibrated structures (both pure DM as well as DM+gas simulations) and that we systematically test the effects of using either spherical or potential bins, of the structures non-sphericity etc.


Combining Eq.~(\ref{eqn:sig}) with Eq.~(\ref{eqn:CJE}) (for a perturbed system, i.e., $\sigma_\phi^2 \rightarrow \tilde{\sigma}_\phi^2$) leaves us
with a Jeans equation containing four terms.  The 3
normal ones, dealing with the density, mass and velocity dispersion
profile, and one new term describing the effect that rotation
has on the system
\begin{equation}
\sigma_r^2 \frac{d\ln (\rho \sigma_r^2)}{d\ln r} + 2\beta\sigma_r^2  -    
\eta \, 12 \, \alpha_4 \, v_{\textrm{rot}} \sqrt{ \sigma_{\phi}^2}  = - \frac{GM}{r}
\label{eq:new} 
\end{equation} 
where the anisotropy is given by $\beta = 1 - \sigma_{\theta}^2 /
\sigma_r^2$. Here we assume that $\sigma_{\theta}^2 =
\sigma_{\phi}^2$, which states that the thermal velocity moments
in the tangential directions are equal, irrespective of the magnitude
of the (small) bulk rotation.
Note that simply substituting $\tilde{\sigma}_\phi^2$ into the (perturbed) Jeans equation leaves us with a Jeans equation, only involving the unperturbed velocity dispersions.

The dominating terms in Eq.~(\ref{eq:new}) are the derivative and the
mass terms. Making the conjecture that the rotation term, which is just a minor perturbation of the Jeans equation, must follow the profile of the dominating mass term, and assuming (for now) that $\beta=0$, we get directly from Eq.~(\ref{eq:new}) a relation between the rotational perturbation and the dominating mass, which reads
\begin{equation}
v_{\textrm{rot}}
\sqrt{\sigma_\phi^2}  \sim \frac{GM}{r} \,.
\label{eq:vm}
\end{equation}
In principle many other solutions, than the conjecture of the small term following the dominant one used above, are allowed to exist, but these would all imply some degree of compensation or fine-tuning between the various terms.
We therefore suspect that there is a more physical explanation
for why  the $v_{\textrm{rot}} \sqrt{\sigma_\phi^2}$ term is proportional to $\frac{GM}{r}$ than our conjecture, but none has been found so far.


The different structures may have fairly different magnitudes of
the angular momentum, and Eq.~(\ref{eq:vm}) expresses only that the
radial {\em profile} of the angular momentum is always the same,
however, the absolute {\em magnitude} is unknown, and may vary from
structure to structure.

In a similar way we can look at the relationship
between the $\beta \neq 0$ and the rotational term. We find from
Eq.~(\ref{eq:new}) that this connection is
\begin{equation}\label{eq:vbeta}
\sigma_r^2 \beta \sim
v_{\textrm{rot}} 
\sqrt{\sigma_\phi^2} \; .
\end{equation}
This relation implies that if $\beta$ goes to 0, the rotation term should go to 0 as well. Since we are here suggesting a relation between the two minor terms in Eq.~(\ref{eq:new}) the relation~(\ref{eq:vbeta}) might not be as strong as relation~(\ref{eq:vm}). The case $\beta < 0$ does not occur in the equilibrated part of the simulated DM halo structure and has therefore no relevance to the problem at hand.

We are aware that $\beta$ is marginally smaller than 0 in DM05. In fact some of our structures also have $\beta<0$ in some of the inner most bins. But since we, as well as DM05, are working with simulations which are known to have difficulties simulating structures at the innermost parts, such values (which are not much below 0) must be considered in agreement with 0 within errors and does therefore not conflict our suggested relation between beta and $\vrot$. If on the other hand cosmological simulations were to produce an equilibrated structure with a clear trend that a fully resolved smooth region of the structure had $\beta < 0$ this would definitely question the validity of our work.

Note that including a centrifugal term into the equations will basically
give a small energy conserving perturbation, which goes as $\vrot^2$,
to the new azimuthal velocity dispersion. However this perturbation is
so small that it is not visible in the numerical simulations, and it
is therefore ignored.

It is now straight forward to test these suggested relations with the
results from numerical simulations.


\section{Comparing with numerical simulations}\label{sec:sim}

We have argued that there may be clear relations between the new
rotational supplement to the Jeans equation and the mass- and
anisotropy-terms. To test this we used 10 intermediately resolved galaxy and cluster sized numerical simulations of  DM halos \citep{mac07}, one high resolution cluster, C$_{HR}$.W3, and one high resolution galaxy, the 'Via Lactea' simulation \citep{diemand,diemand2}.

The 10 intermediately resolved simulations  have been  performed using  PKDGRAV,  a treecode
written by Joachim Stadel and Thomas Quinn \citep{stad01}.  The initial
conditions are generated with the GRAFIC2 package \citep{bert01}.
The starting  redshifts $z_i$  are set to  the time when  the standard
deviation  of the  smallest density  fluctuations resolved  within the
simulation box  reaches $0.2$ (the smallest scale  resolved within the
initial conditions is defined as twice the intra-particle distance).
All the halos were identified using a SO (Spherical Overdensity) 
algorithm  \citep{mac07}.
The cluster-like haloes have been extracted from a 63.9 $Mpc/h$
simulation containing $600^3$ particles, with a mass resolution of 
$m_p=8.98 \times 10^7 M_{\odot}/h$. The masses of the clusters used for 
this study are 2.1, 1.8, and 1.6 $\times 10^{14} M_{\odot}/h$.
The galaxy sized haloes have been obtained by re-simulating at high resolution 
haloes found in the previous simulation. The simulated haloes 
are in the mass range $0.9-2.5 \times 10^{12} M_{\odot}/h$ and have a mass
resolution of $m_p=4.16 \times 10^5 M_{\odot}/h$ that gives a minimum number of particles
per halo of about $2.5 \times 10^6$ particles.
The High resolution cluster C$_{HR}$.W3, based on the PKDGRAV as well,  has 11 millions particles within its virial radius
and a mass of $ M=1.81\times 10^{14} M_{\odot}/h$.
The  'Via Lactea' (which is also based on the PKDGRAV code)
simulation includes 234 million particles with force resolution of 90
pc, and it includes one highly equilibrated structure of mass $M_{200}
= 1.77 \times 10^{12} M_\odot$, containing about 84 million particles \citep{diemand}.

 \begin{figure}
 \epsscale{0.95}
 \plotone{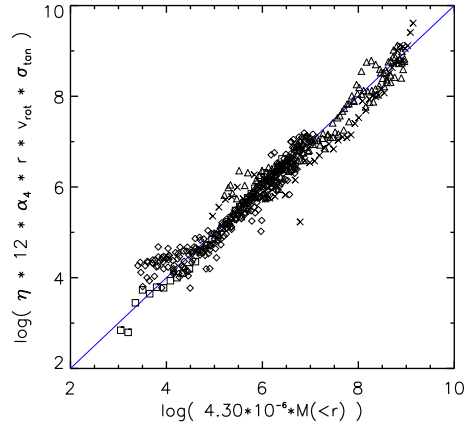}
	\caption{The suggested connection between the mass (r.h.s.\ of
	the Jeans equation) on the x-axis, and the new rotational term (solid
	line) on the y-axis. The diamonds, triangles, crosses and the squares represent the galaxy sized halos and the cluster sized halos from the simulations \cite{mac07}, the high resolution C$_{HR}$.W3 cluster and the \cite{diemand} "Via Lactea" high resolution simulation respectively. This illustrates that the Jeans
	equation determines the radial behavior of the rotation,
	i.e. the angular momentum $j(r)$. We use SI units
	and $\alpha_4=3$.  The factor $4.30\cdot10^{-6}$
	includes the gravitational constant $G$, and is the correction
	needed to have the quantities in SI units. The $\eta$ for each structure corresponds to the ones listed in Table~\ref{tab:eta}. }
\label{fig:vmall}
\end{figure}

Plotting the rotation-term against the mass for the different simulations gives Fig.~\ref{fig:vmall}. Here the diamonds, triangles, crosses and squares represent the galaxy sized halos, the cluster sized halos, the C$_{HR}$.W3 simulation and the \cite{diemand} "Via Lactea" high resolution simulation respectively.
In Fig.~\ref{fig:vmall} we see a clear linear relation between the two
terms. This means that the generalized Jeans equation
(Eq.~(\ref{eq:new})) determines the radial behavior of the rotation,
i.e. the angular momentum $j(r)$. This also explains why
\cite{bullock} find a strong relation between the angular momentum and
the mass in their simulations, since our conjecture resembles the
results of \cite{bullock} when $\sigma_\phi^2$ is constant.
We have tested that this relation is not just an effect of choosing (actually deriving) a term in the Jeans equation with the right units. For instance the term with $v_{\textrm{rot}}^2 r$ does not have a correct relation to the mass (as \cite{bach} also conclude).

\begin{figure}
 \epsscale{0.95}
 \plotone{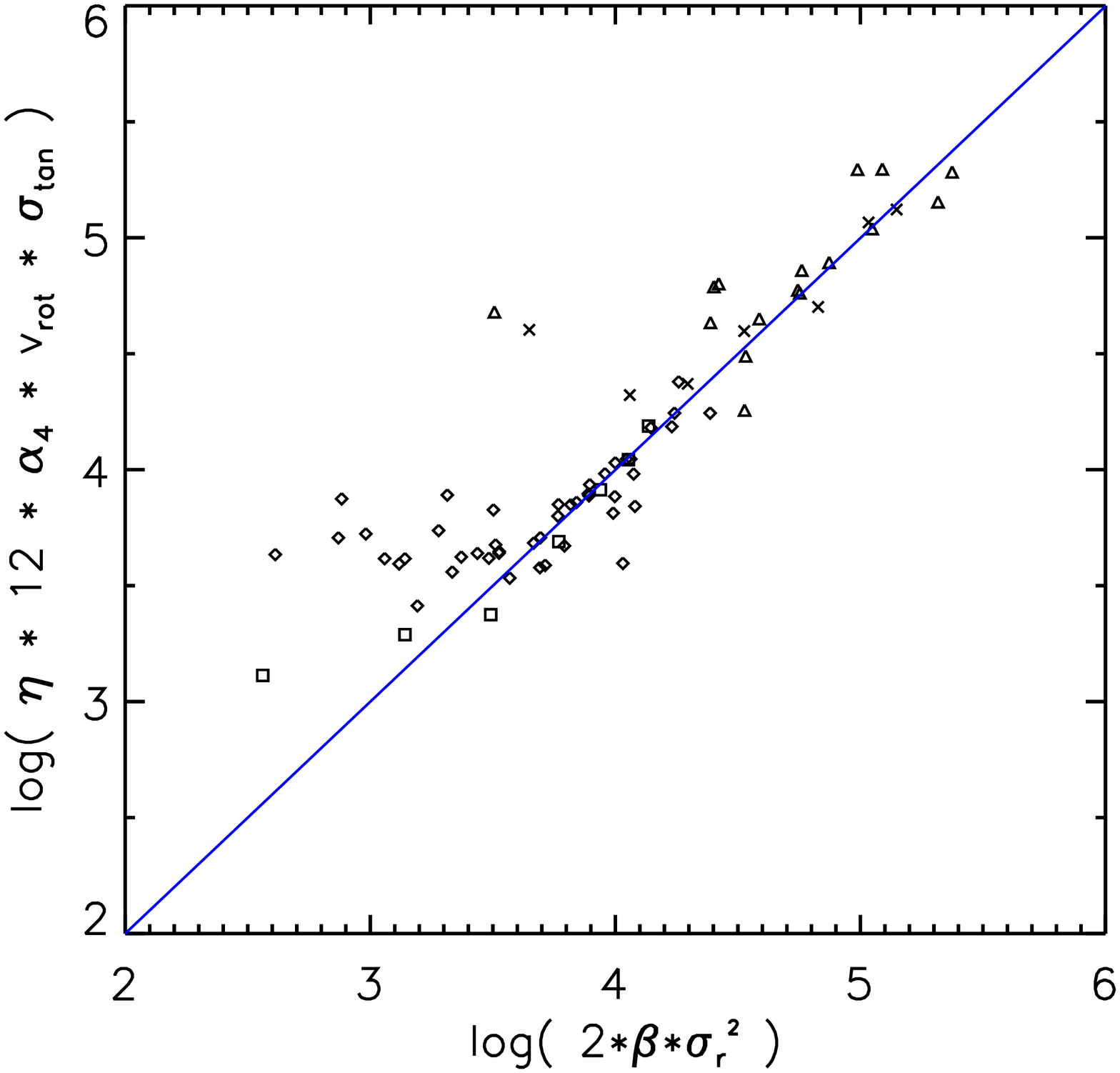}
	\caption{The relation between the thermal velocity anisotropy
	$\beta$ and the rotation-term (straight line). The squares
	are from the high resolution numerical simulation of
	\cite{diemand} and the diamonds are the galaxy sized halos from the \cite{mac07} simulations. The crosses is the C$_{HR}$.W3 high resolution cluster and the triangles are the cluster sized structures from \cite{mac07}. We see an indication of a deviation from the relation for small $\beta$. This is probably a consequence of comparing the two subdominant terms in the new Jeans equation. When plotting the data we have cut off the structures so that all of the points can be considered to be equilibrated. We have determined the cutoff-points by excluding the (outer) part of the structures, where the velocity anisotropy is not a (roughly) monotonically increasing function of radius. Furthermore we have re-binned the data to reduce scatter, so that each point now contains roughly 1/7 of the structure. We again use SI units, $\alpha_4=3$ and different $\eta$ for each structure as listed in Table~\ref{tab:eta}. }
\label{fig:vbeta}
\end{figure}

In a similar way we can test our suggested linear relation
between $\sigma_r^2 \beta$ and $v_{\textrm{rot}}
\sqrt{\sigma_{\phi}^2}$. Plotting these quantities for the intermediate resolution halos together with the C$_{HR}$.W3 and 'Via Lactea' high resolution simulations gives
Fig.~\ref{fig:vbeta}. Here we see a clear
correlation for the majority of $\beta$ values. However, there is some indication that small $\beta$ values doesn't follow our relation as strictly as for large $\beta$ values. This is probably a consequence of comparing the two subdominant terms in the new Jeans equation with one another, which as mentioned doesn't make relation (\ref{eq:vbeta}) as strong as the relation between the mass and the bulk rotation. 
In the figure we have re-binned the data to reduce scatter and cut off the structures where $\beta$ was no longer a (roughly) monotonically increasing function of radius. Plotting the structures without making an outer cut doesn't change the picture but only enhances the overall scatter. 

\begin{deluxetable*}{|l|c|c|c|c|c|c|c|}
\hline
Simulation & $\eta_\beta$ & $d\eta_\beta$ & $\eta_M$ & $d\eta_M$ & $r_{\mathrm{vir}} \; [kpc] $ & $M_{\mathrm{vir}} \; [M_\odot] $ & $\sqrt{\sigma_{\phi\mathrm{, vir}}^2} \; [km/s] $  \\
\hline
'Via Lactea' 	& 0.26 & 0.05 & 0.32 & 0.15 & 359 	& 5.98e+11 & 89.1 \\
G0.W1 		& 0.25 & 0.10 & 0.70 & 0.40 & 260 	& 1.24e+12 & 79.1 \\
G1.W1 		& 0.20 & 0.10 & 0.60 & 0.30 & 288 	& 1.54e+12 & 97.7 \\
G1.W3 		& 0.28 & 0.08 & 0.65 & 0.33 & 333 	& 1.89e+12 & 77.6 \\
G2.W1 		& 0.17 & 0.04 & 0.45 & 0.20 & 339  	& 2.63e+12 & 128 \\
G2.W3 		& 0.32 & 0.08 & 0.80 & 0.35 & 288 	& 1.11e+12 & 61.2 \\
G3.W1 		& 0.22 & 0.05 & 0.63 & 0.25 & 296 	& 1.76e+12 & 94.4 \\
G4.W3 		& 0.31 & 0.07 & 0.85 & 0.45 & 218 	& 5.96e+11 & 59.0 \\
C1.W3 		& 0.07 & 0.04 & 0.22 & 0.15 & 1440 	& 2.13e+14 & 550 \\
C2.W1 		& 0.07 & 0.04 & 0.23 & 0.10 & 1440 	& 2.12e+14 & 484 \\
C3.W1 		& 0.18 & 0.07 & 0.45 & 0.25 & 1600 	& 1.89e+14 & 444 \\
C$_{HR}$.W3 	& 0.24 & 0.06 & 0.80 & 0.60 & 1671 	& 3.23e+14 & 557 \\
\hline
\caption{Our free parameter $\eta$ from Figs.~\ref{fig:vmall} and \ref{fig:vbeta}, its errors and the mass and azimuthal velocity dispersion taken at the virial radius $r_{\mathrm{vir}}$, for the structures used to test our relations.  The errors on $\eta_\beta$ and $\eta_{\mathrm{mass}}$ represent the interval in which the values give the best fit to the relations. For instance the 'Via Lactea' fits the mass relation in Fig.~\ref{fig:vmall} reasonably well for $ 0.17 < \eta_{\mathrm{mass}} < 0.47$, i.e., $0.32\pm0.15$. All the structures are based on the PKDGRAV tree code. The 'Via Lactea' and the GX.XX are galaxy sized structures and the C$_{HR}$.W3 and CX.XX are cluster sized structures. The XX.W1 is based on a WMAP 1-year data cosmology ($h=0.71$ and $\Omega_m = 0.268$), whereas XX.W3 is based on a WMAP 3-year data cosmology ($h=0.73$ and $\Omega_m = 0.238$). For more information on the structures, refer to the text or see \cite{diemand} and \cite{mac07}.}
\label{tab:eta}
\end{deluxetable*}

We see that the rotation term goes to 0 as $\beta$ goes to 0, exactly as suggested. 
In fact, if we plot the fraction of the kinetic energy in rotation, i.e., $\vrot^2/\sigma_\phi^2$, we see that it drops from $10^{-2}$ in the outskirts of the structure, down below $10^{-4}$ for the inner-most bins. Since $\beta$ is monotonically increasing as a function of radius, the fact that the rotation becomes so small in the inner parts of the structure supports our suggested relation of the rotation term going towards 0 for small $\beta$.

In Figs.~\ref{fig:vmall} and \ref{fig:vbeta} the only free parameter in our relations, $\eta$, have been fitted for each structure. These values of $\eta$ corresponding to the relations in Eqs.~(\ref{eq:vm}) and (\ref{eq:vbeta}) represents the unknown magnitude of the angular momentum and are shown together with the estimated errors in Table~\ref{tab:eta}.

\begin{figure}
 \epsscale{0.95}
 \plotone{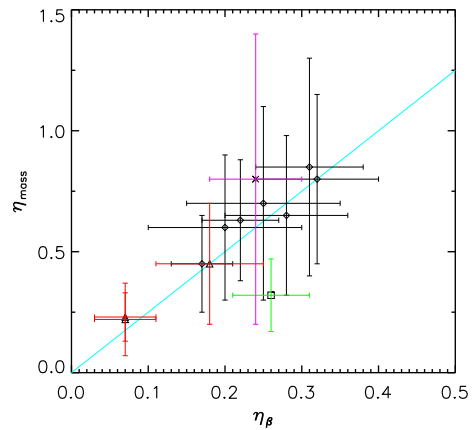}
	\caption{The free parameter $\eta_{\textrm{mass}}$ obtained from the relation plotted in Fig.~\ref{fig:vmall} and $\eta_\beta$ obtained from the relation in Fig.~\ref{fig:vbeta}, plotted against each other. 
The solid guide-the-eye line represents the relation $\eta_{\textrm{mass}} = 2.5\eta_\beta$. 
The error bars represent the interval in which the $\eta$ values give the best fit to the relation. For instance the 'Via Lactea' (square) fits the mass relation in Fig.~\ref{fig:vmall} reasonably well for $ 0.17 < \eta_{\mathrm{mass}} < 0.47$, i.e., $0.32\pm0.15$ as written in Table.~\ref{tab:eta}. The symbols are the same as in Fig.~\ref{fig:vmall}.}
\label{fig:etaeta}
\end{figure}

\begin{figure}
 \epsscale{0.95}
 \plotone{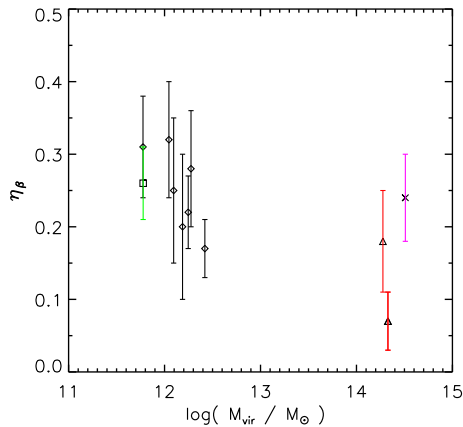}
	\caption{The free parameter $\eta_\beta$ plotted against the virial mass of each structure (listed in Table \ref{tab:eta}. The symbols are the same as in Fig.~\ref{fig:vmall}.}
\label{fig:etam}
\end{figure}

Plotting the $\eta$ values and their errors gives Fig.~\ref{fig:etaeta}. Here we see a tendency of $\eta_{\mathrm{mass}}$ being larger than $\eta_\beta$. Since the triangles and the cross are cluster like structures and the rest are galaxy like structures we notice that there might be a connection between $\eta$ and the mass of the structures. In Fig.~\ref{fig:etam} we plot $\eta_\beta$ (since it has the smallest error bars, percentage-wise) against the estimated virial mass of each structure (see Table~\ref{tab:eta}), and see that our free parameter anti-correlates slightly with the mass of the structure. So according to our suggested relations the effect an added bulk rotation has on a system is anti-correlated with the virial mass of that system.

\begin{figure}
 \epsscale{0.95}
 \plotone{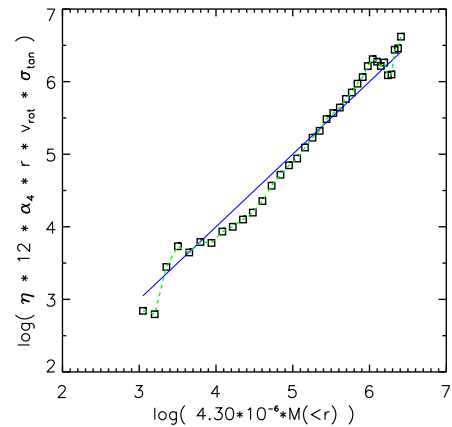}
	\caption{The same plot as Fig.~\ref{fig:vmall} containing only the data from the \cite{diemand} "Via Lactea" high resolution simulation . We only plot numerically
	simulated points (squares) in the resolved region,
	i.e. outside $r=1$ kpc, and out to $r=r_{200}$.  The outermost
	6 to 10 points (top right corner) are possibly not yet fully equilibrated (as can be seen when looking at the profile of $\frac{d\ln\rho}{d\ln r}$) and might therefore be ignored. We use SI units,
	$\eta=0.32$ and $\alpha_4=3$.  The factor $4.30\cdot10^{-6}$
	includes the gravitational constant $G$, and is the correction
	needed to have the quantities in SI units.}
\label{fig:vM}
\end{figure}

\begin{figure}
 \epsscale{0.95}
 \plotone{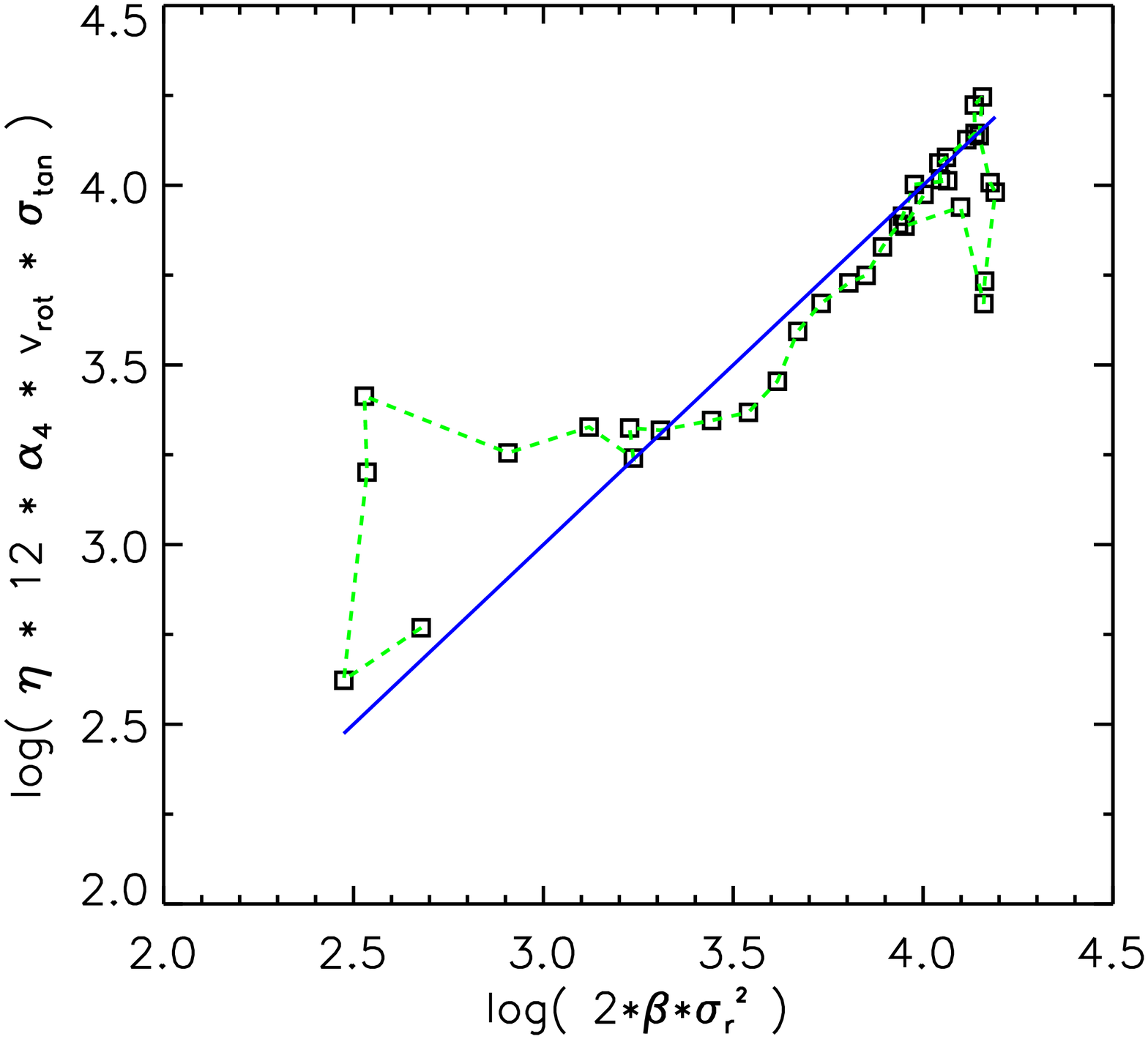}
	\caption{The relation between the thermal velocity anisotropy
	$\beta$ and the rotation-term (straight line), and the high resolution numerical simulation of
	\cite{diemand} (squares).  The trend that lower $\beta$ doesn't obey our relation as well as large $\beta$ is also obvious here. We again use SI units, $\eta=0.26$ and
	$\alpha_4=3$.  The outermost 6 to 10 points are possibly not
	yet fully equilibrated (clearly visible when considering $\frac{d\ln\rho}{d\ln r}$ as a function of radius) and might therefore be ignored.}
\label{fig:vbetadiem}
\end{figure}

After having tested our suggested relations from the previous section with the simulations from \cite{mac07}, we also held them up against the recent high resolution
numerical simulation 'Via Lactea' by \cite{diemand}, to make sure that the results is not just a coincidence because of lack of numerical resolution. 
We have plotted the high resolution data as squares in Figs.~\ref{fig:vmall} to \ref{fig:etam} for comparison. In Figs.~\ref{fig:vM} and \ref{fig:vbetadiem} we have plotted the 'Via Lactea' alone, without any cutoffs or re-binning. In these figures we are using the values $\eta$ from Table ~\ref{tab:eta}. On both figures we see that the suggested relations are confirmed when comparing with highly resolved data. 
The 'Via Lactea' structure did not experience any major mergers since $z=1$.  All quantities are extracted in spherical bins. Due
to numerical softening one can safely trust the radius outside 1 kpc  of this galaxy. In this simulation the
outermost 6-10 points should be considered with care since they are
potentially not fully equilibrated yet, as is easily seen when considering
the radial derivative of the density profile.

We have thus compared the angular momentum of the intermediate resolution structures, the highly resolved C$_{HR}$.W3 cluster, and the 'Via Lactea' simulation, which is one of the best resolved structures published today~\citep{diemand}, to our suggested relations, and see strong correlations between the rotation, mass and velocity anisotropy of the system.

\begin{figure}
 \epsscale{0.95}
 \plotone{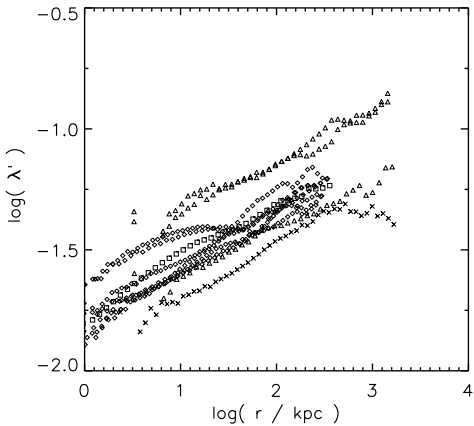}
	\caption{The spin parameter (Eq.~(\ref{eq:lam})) plotted as a function of radius. The symbols are the same as in Fig.~\ref{fig:vmall}.}
\label{fig:lambdar}
\end{figure}

As mentioned in the introduction, people often use the spin parameter $\lambda$ when describing the angular momentum of DM halos. Combining our relation between the mass and the angular momentum with the spin parameter, as defined by \cite{bullock}
\begin{equation}
\lambda' = \frac{J}{\sqrt{2GM^3r}}
\end{equation}
we end up with a new expression for the spin parameter
\begin{equation}\label{eq:lam}
\lambda' = \frac{1}{\sqrt{2}\;\eta \, 12 \, \alpha_4} \; \frac{v_c}{\sqrt{ \sigma_{\phi}^2} }
\end{equation}
where $v_c^2 = GM/r$. We are therefore able to describe the spin parameter only as a function of mass and $\sqrt{\sigma_\phi^2}$, without any dependence on the bulk rotation $\vrot$. In Fig.~\ref{fig:lambdar} we have plotted Eq.~(\ref{eq:lam}) for the simulated structures. This figure agrees fairly well with Fig.~5 of \cite{ascasibar}. If we estimate a gradient of our plot we get approximately 1/4 to 1/5 (depending on the chosen structure), whereas an estimated gradient on Fig.~5 in \cite{ascasibar} is closer to 1/6. Thus Eq.~(\ref{eq:vm}) appears to roughly explain the observed tendency of an increase in the spin parameter as a function of radius.
As mentioned earlier a relation with a $r \vrot^2$ term instead of the $r \vrot \sqrt{\sigma_\phi^2}$  term we suggest will according to simulations not give the correct relation to the mass. Furthermore a result on the $r \vrot^2$ form implies a constant spin parameter as a function of radius, and since this is clearly not in agreement with the work by \cite{ascasibar}, we take this as yet another indication of the success of the relations in Eqs.~(\ref{eq:vm}) and (\ref{eq:vbeta}). 

\begin{figure}
 \epsscale{0.95}
 \plotone{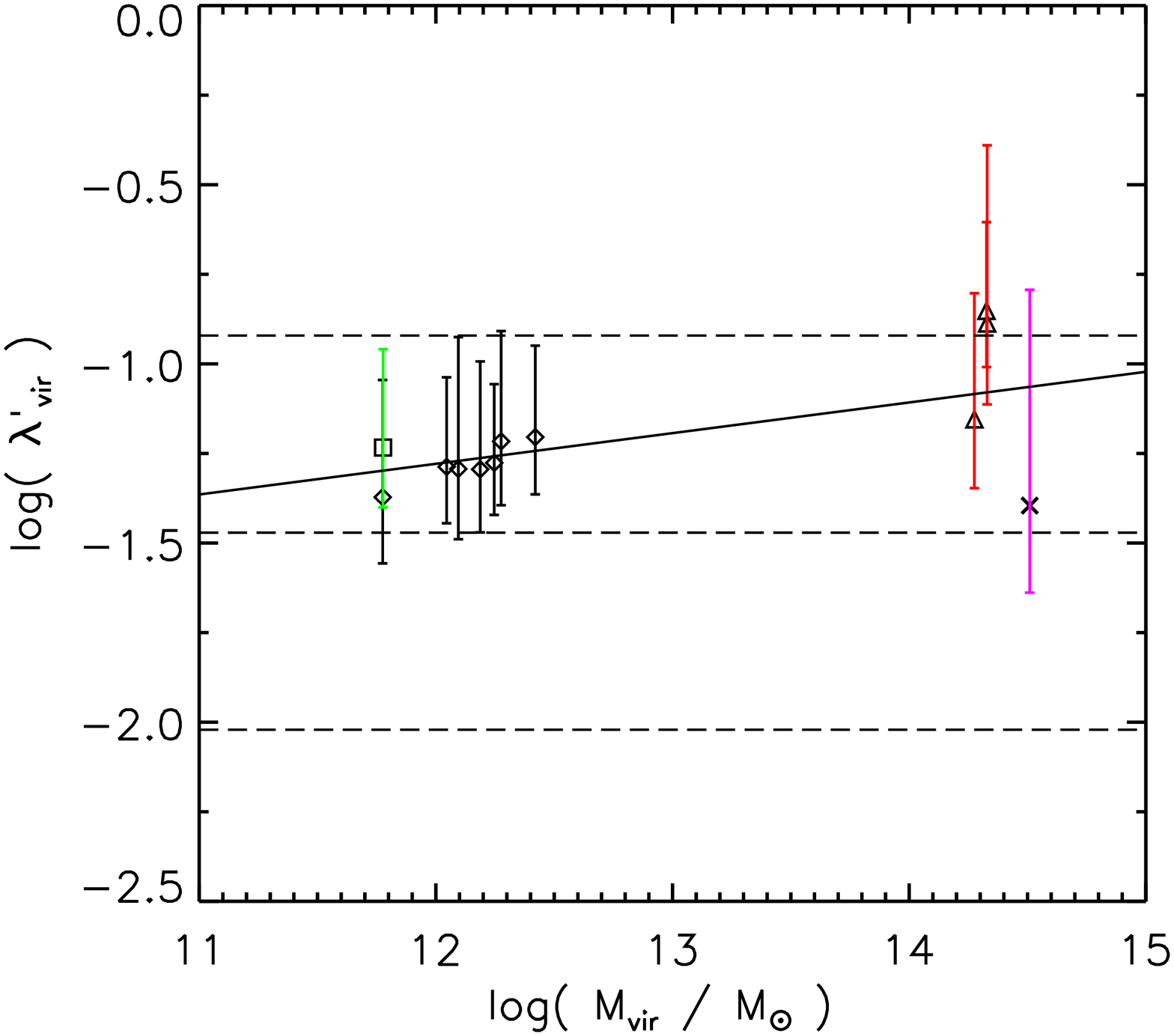}
	\caption{The spin parameter (Eq.~(\ref{eq:lam})) at the virial radius. The linear fit to the points (straight line) is given by $(0.09 \pm 0.04) \times \log(M_{\mathrm{vir}}/M_\odot) - (2.31 \pm 0.51)$. The errors correspond to the error in $\eta_{\textrm{mass}}$ (see table \ref{tab:eta}) used when calculating $\lambda'_{\mathrm{vir}}$. The dotted lines represents the mean ($0.034 \pm 0.001$) and $2\sigma$  scatter ($0.55 \pm 0.01$) from \cite{mac07}, which are in agreement with the results of \cite{bullock}. The symbols are the same as in Fig.~\ref{fig:vmall}. }
\label{fig:lambdam}
\end{figure}

Finally it have been suggested that the spin parameter doesn't depend on the virial mass of the structures \citep{mac07}. To test this we plot in Fig.~\ref{fig:lambdam}, the values of $\lambda'$ from Eq.~(\ref{eq:lam}) taken at the virial radius of the structures. To calculate $\lambda'_{\textrm{vir}}$ we used the values given in Table~\ref{tab:eta}. Here we see the indication of a slight increase in spin as the virial mass of the structures grow. The linear fit in Fig.~\ref{fig:lambdam} (full line) has an inclination of $0.09 \pm 0.04$.  One possible explanation for the indication of a mass dependence, might be the fact that the spin is taken at $r_{\mathrm{vir}}$. As pointed out by \cite{ascasibar} the use of  $r_{\mathrm{vir}}$ (compared to their $R_{\mathrm{max}}$) might be too 'non-conservative' when estimating the various properties of equilibrated structures.
On the other hand it is not surprising with a slight increase in spin, since the spin parameter as defined in Eq.~(\ref{eq:lam}) basically resembles a relation between $\eta_{\textrm{mass}}$ and $M_{\textrm{vir}}$ similar to the one shown for $\eta_\beta$ in Fig.~\ref{fig:etam}.
Nevertheless, because of the large scatter and error-bars in our points we must conclude that there is no (significant) dependence between our spin parameter and the virial mass of the structures. This is in agreement with the lower part of Fig.~3 in \cite{mac07}, which also shows that a relatively large scatter in the points is usual.

\section{Conclusions}

We have studied the form of the spherical Jeans equation when one
includes angular momentum, and we find that a bulk motion leads to the
introduction of an extra term, which includes the average rotational
velocity. This is done under the assumption that the distortion of the distribution function takes the form argued for in Eq.~(\ref{eqn:ff}). This assumption is supported by numerically simulated structures.
The distortion enables us to suggest a new correlation between the angular momentum and the mass.
This relation
is in good agreement with the findings of recent high resolution
numerical simulations of cosmological structures. We also suggest a
correlation between the angular momentum and the velocity anisotropy
profiles, which is also in fair agreement with numerical
findings. 
These suggested relations imply that cosmological dark matter
structures have angular momentum profiles which have the same
universal properties, irrespective of how or when they were formed.

Finally we derive a new form of the spin parameter, $\lambda'$, which is shown to increase slowly as a function of radius, in agreement with recent simulations. Furthermore our relations indicate that there is no (significant) dependence between $\lambda'_{\mathrm{vir}}$ and $M_{\textrm{vir}}$.

\section*{Acknowledgment}

It is a pleasure to thank Juerg Diemand and Jesper Sommer-Larsen for kindly providing the
numerically simulated data used in the figures. We thank the anonymous referee for suggestions which significantly improved the paper.
This work was initiated during the 
``{\em Dark Matter Workshop}'' in Copenhagen, organized by
``{\em Niels Bohr International Academy}'' and 
``{\em Dark Cosmology Centre}''.  
LLRW would like to acknowledge NSF grant AST-0307604 which allowed her
to attend the Copenhagen workshop.
Part of the numerical simulations were performed on  the  PIA 
cluster  of  the
Max-Planck-Institut f\"ur Astronomie at the Rechenzentrum in Garching.
The Dark Cosmology Centre is 
funded by the Danish National Research Foundation.


\end{document}